Temperature evolution of structural and magnetic properties of stoichiometric LiCu$_2$O$_2$: correlation of thermal expansion coefficient and magnetic order.

S. A. Ivanov[1,2], P. Anil Kumar[1], R. Mathieu[1,*], A. A. Bush[3], M. Ottosson[4], P. Nordblad[1]

[1] Department of Engineering Sciences, Uppsala University, Box 534, SE-751 21 Uppsala, Sweden
[2] Center of Materials Science, Department of Inorganic Materials, Karpov Institute of Physical Chemistry, Moscow, 105064, Russia
[3] Moscow State Technical University of Radio Engineering Electronics, and Automation, pr. Vernadskogo 78, Moscow, 119454 Russia
[4] Department of Chemistry, Uppsala University, Box 538 SE 75121, Uppsala, Sweden.

*Tel.: +46-(0)18 471 72 33; Fax: +46-(0)18 471 32 70; email address: roland.mathieu@angstrom.uu.se

**Abstract**

Temperature-dependent crystallographic and magnetic studies on stoichiometric single crystals of LiCu$_2$O$_2$ are reported. The magnetic properties are similar to earlier findings demonstrating antiferromagnetic ordering below 25 K. Evidence of magnetoelastic coupling is observed in the thermal expansion along the *c*-direction; not only at the low temperature antiferromagnetic transitions, but an anomalous behavior of the thermal expansion indicate magnetoelastic coupling also to the magnetic ordering related to a weak spontaneous magnetic moment appearing at 150 K. Ac-susceptibility measurements at different frequencies and superposed dc-fields are employed to further characterize this magnetic anomaly.

**Introduction**

Low-dimensional quantum spin systems with geometric frustration have attracted much attention during the last decades. Compounds with edge-sharing CuO$_2$ chains and Cu-O-Cu bond angle near 90° such as LiCu$_2$O$_2$ are frustrated magnets owing to competition between ferro- and antiferromagnetic exchanges and exhibit rich sets of magnetic phases [1]. The low dimensional nature of LiCu$_2$O$_2$ and the associated quantum spin fluctuations bring forth intriguing magnetic and electronic properties. For example, LiCu$_2$O$_2$ belongs to a family of magnetoelectric cuprates [1] in which complex magnetic order induces electric polarization, which can be reversibly flipped with an applied magnetic field [2-4]. LiCu$_2$O$_2$ orders magnetically at low temperature via two successive transitions (about 24 and 22 K) to a "polar" helical magnetic structure [2-4]. The exact nature and origin of this magnetic structure (e.g. propagation direction and planes) is still under debate [2-12].
The crystal structure of LiCu$_2$O$_2$ has orthorhombic symmetry (space group *Pnma* with $a$ = 5.730(1), $b$ = 2.8606(4), $c$ = 12.417(2) Å) and represents an alternation of three layers along the *c* axis: (**i**) –Cu$^+$(1)–, (**ii**) –O(1)Cu$^{2+}$(2)O(2)Li–, and (**iii**) –LiO(2)Cu$^{2+}$(2)O(1)– [13]. As displayed in Fig. 1, chains of Cu$^{2+}$ ions propagate along the *b*-axis. The magnetic structure of LiCu$_2$O$_2$ is made up of these exchange-coupled double linear Cu$^{2+}$ chains (two-leg ladder system), which belong to two neighboring LiCuO$_2$ layers. Cu$^{2+}$ ions in the ladder systems form interacting Heisenberg spin-1/2 chains. In an ideal ladder structure, no long-range magnetic order should be realized. The antiferromagnetic ordering observed in LiCu$_2$O$_2$ has been suggested to originate from a partial intermix of copper and lithium ions, which gives rise to exchange interaction between the isolated ladders through the Cu$^{2+}$ ions incorporated into the Li chains [6].
Crystals of LiCu$_2$O$_2$ contain univalent (Cu$^+$) and divalent (Cu$^{2+}$) copper cations of equal amount distributed over their structural positions in an ordered manner [13]. According to the

Goodenough-Kanamori rules [14], the exchange interaction between the nearest-neighbor $Cu^{2+}$ spins is weak and ferromagnetic (FM) in nature while the next-nearest-neighbor interaction is antiferromagnetic (AFM). Such competing interactions can lead to an incommensurate helimagnetic structure and in $LiCu_2O_2$ classical helical magnetism has been revealed in neutron experiments [5,6]. The magnetic and magnetoelectric properties of $LiCu_2O_2$ depends on Li, $Cu^+$, $Cu^{2+}$ arrangement. Interestingly, it was recently suggested that cooling conditions through the structural high-temperature transition (993 K) in $LiCu_2O_2$ could induce Li-Cu cation redistribution, thereby affecting magnetic properties and thus polarization [15]. The magnetic interaction may also be affected by anion or cation excess or deficiency. For example, $Cu^{2+}$ impurity spins created by Li deficiencies were found to affect the low-temperature helical magnetic structure of the system [16]. Low temperature investigations of the magnetic properties of non-twinning $LiCu_2O_{2+\delta}$ crystals [17] showed that a weak spontaneous magnetization arises in the compound at about 150 K much above the ordering temperature of the basic long-range helimagnet phase at $T<24$ K.

In this article we report results on the temperature dependence of crystallographic and magnetic properties on stoichiometric single crystal of $LiCu_2O_2$. The lattice thermal expansion along the *c*-direction anomalously changes slope at about 150 K where also the weak spontaneous magnetic moment appears. This suggests correlation between the structure and the magnetic order related to the spontaneous magnetic moment along the c-axis.

**Experimental**

Stoichiometric single crystals of the $LiCu_2O_2$ were grown by a flux method under slow cooling of the melt of a $20Li_2CO_3 \cdot 80CuO$ mixture in alundum crucibles in air following method described in details in [15]. The mixture of analytical grade $Li_2CO_3$ and CuO composition was melted at 1100 °C, and then solidified by cooling to 930 °C at a rate of 5.0 °C/hour. Next, the crucible was withdrawn from the furnace and placed on a massive copper plate to ensure rapid cooling to room temperature. Quenching from about 900 °C was necessary, because the single phase $LiCu_2O_2$ single crystals decompose below this temperature [15,18]. The crystals were shaped as flat plates developed parallel to the *ab* plane. In most cases, the samples were twinned with a characteristic domain size of several microns. It was possible to select samples without twinning structure as confirmed by x-ray diffraction and ESR measurements. Also, thermogravimetric investigations carried out on a Paulik–Erdei Q 1500D computer assisted derivatograph confirmed stoichiometric composition of selected crystals.
X-ray powder diffraction (XRPD) studies were performed on selected single crystals of $LiCu_2O_2$, which were crushed into powder in an agate mortar. Low temperature x-ray diffraction studies were performed on a Si substrate (as a sample holder), which was covered with several drops of the resulting powder suspension of $LiCu_2O_2$ in ethanol, leaving randomly oriented crystallites after drying. XRPD data were collected on Bruker D8 Advance diffractometer (Ge monochromatized Cu K$\alpha$1 radiation, Bragg-Brentano geometry, a Våntec position sensitive detector, DIFFRAC plus software) in the 2θ range 10 – 140° with a step size of 0.02° (counting time was 15 s per step). The slit system was selected to ensure that the x-ray beam was completely within the sample for all 2θ angles. Low temperature diffraction patterns were registered using a closed cycle helium cryostat Phoenix (Oxford Instruments). Control, monitoring and data logging during the measurements were available via Cryopad software. Tilting sample was used during data collection in order to restrict a possible influence of texture. Stability of temperature was about 0.1 K. An internal silicon standard was used for accurate unit cell determination. The sample was cooled and warmed between 12 and 300 K at a rate of 10 K per hr. Data were collected over a 2θ range of 10-140°, giving an average of one scan every 10 K. Unit cell parameters as a function of temperature were extracted from data analysis using the Bruker DIFFRAC software package. An orthorhombic structural model for $LiCu_2O_2$ in space group *Pnma* was used throughout the temperature range observed.

The dc magnetization and ac-susceptibility of a LiCu$_2$O$_2$ single-crystal was recorded with the dc and ac magnetic fields applied along the *c*-axis of the structure on an MPMS SQUID magnetometer from Quantum Design Inc.

**Results and Discussion**

Our stoichiometric single crystals of LiCu$_2$O$_2$ display a low-temperature magnetic behavior similar to that reported earlier [2-4]. Fig. 2 (a) shows zero-field-cooled (ZFC) and field-cooled (FC) magnetization curves measured in a field of 1 kOe applied along the *c*-axis of the crystal as a function of temperature. The low-temperature downturn of the magnetization below 30K associated with the complex magnetic ordering is evident (the antiferromagnetic transitions are seen more clearly in the temperature-derivative of the magnetization, see refs. [2-4, 8]). A weak anomaly is detected in the magnetization curves near 150 K, below which irreversibility between the ZFC and FC curves is observed. This irreversibility corresponds to the anomaly observed in low field (H = 10 Oe) measurements on LiCu$_2$O$_{2+\delta}$ crystals in the same temperature range [17]. However, as seen in the inset, the magnetization varies essentially linearly with magnetic field in that temperature range. Yet, probing the system with small magnetic fields, as in Ref. [17], a peak can be observed near 150 K in ac-susceptibility curve; see Fig. 2(b). Figs. 2(c) and (d) show respectively the real ($\chi'$) and imaginary ($\chi''$) parts of ac susceptibility for different frequencies (lines with symbols). Both the real and the imaginary parts of the susceptibility exhibit pronounced peaks at 150 K suggesting some magnetic ordering below this temperature. Furthermore, the peaks in the susceptibility are completely suppressed when a large dc bias field (1 or 10 kOe) is superimposed during the ac measurement (see Figs. 2(c) and (d), lines without symbols). Hence the spontaneous moment reported in [17] is also present in our stoichiometric single crystals. It is interesting to note that we have observed a similar magnetic behavior near 150 K in single crystalline platelets grown from torch-flame heating Li$_2$CO$_3$ and CuO mixtures [13]. Although several studies have investigated the low-temperature magnetic state of LiCu$_2$O$_2$ in detail and in the light of the possible frustration of the magnetic interaction, none of them has reported a Curie-Weiss behavior in the system. The ratio between Curie-Weiss temperature and antiferromagnetic transition temperature tells about the degree of frustration in the system [19]. In the present crystals, our *H/M* data suggests a Curie-Weiss temperature of about -100 K, which is significantly larger than. However, as discussed in the introduction, the low-dimensional character of the interaction also impedes magnetic ordering and leaves the weak interlayer coupling as a cause of long range order.

XRPD patterns collected at different temperatures are shown in Figure 3. The phase purity of the powder sample was checked from these patterns, which show the expected set of reflections without any impurity lines. All reflections in the room temperature XRPD data were compatible with a single-phase orthorhombic structure (*Pnma* space group) with *a* = 5.7303(4), *b* = 2.8594(3), *c* = 12.4192(5) Å. These lattice parameters for LiCu$_2$O$_2$ at room temperature are in agreement with those reported in Ref. [13], as quoted above. Splitting of the (400), and the (020) reflections was registered at all temperatures. This splitting could be used as direct evidence that the crystal structure of LiCu$_2$O$_2$ is orthorhombic. Diffraction peaks with the indices h and k different from 0 are found to be broader than those with h=0, k=0. This anisotropic line-broadening is an indication of atomic disorder in the crystallographic (*a,b*)-plane. The in-plane correlation length, as calculated from the half-width at half maximum of the Bragg peaks, amounts to about 600 Å. Both cooling and warming data were registered, showing no thermal hysteresis.

The temperature dependence of the lattice parameters, calculated from the obtained XRPD data (cooling mode) measured at different temperatures, are shown in Figs. 4 (a) and (b). Due to the proximity between and overlap of in-plane reflections, as exemplified in the inset of Fig. 3, only the $c$-axis parameter could be extracted independently. Anomalous changes of the $a$- and $c$-lattice parameters were registered around the magnetic phase transitions at about 25 K. There are no deviations from orthorhombic symmetry or any dramatic changes in the behavior of the lattice parameters that suggest additional structural phase changes in the sample at low temperatures. Yet, recent helium-atom scattering studies suggest surface structural rearrangements in same temperature range [20]. The orthorhombic strain $(a-2b)/(a+2b)$ is remarkably found to increase with temperature as shown in Fig. 4 (c) (from $0.47 \times 10^{-3}$ at 12 K to $1.01 \times 10^{-3}$ at room temperature, akin to the results presented in Ref. [21]. This is quite unusual because for most materials there is a reduction of the lattice strain with increasing of temperature.

The linear thermal expansion coefficients defined as $\alpha_x = 1/x(dx/dT)$, $x = a, b, c$ were locally derived from the T-dependent lattice parameter data. No significant feature was observed near 150 K in $\alpha_a$ and $\alpha_b$. However, apart from the low temperature decrease with temperature associated with the antiferromagnetic ordering around 25 K, $\alpha_c$ shows, as seen in Fig. 4d, an anomalous crossover from a near linear increase below 150 K to a weakly temperature dependent behavior at higher temperatures. This crossover temperature coincides with the temperature where a weak spontaneous magnetization appears in the magnetic measurements. Looking closer at Fig. 4d and comparing the expectations on the temperature dependence of the thermal expansion of a non-magnetic crystal to what is observed along the $c$-axis of $LiCu_2O_2$ – an anomalous decay is observed at low temperatures that can be assigned to magnetoelastic coupling and the magnetic transitions at 22 and 24 K. At higher temperatures, where a slow approach to normal behavior could be expected, we instead observe a superposed anomalous bumpy signature centered around 150 K. This feature can be assigned magnetoelastic coupling to the magnetic order revealed by the occurrence of the weak spontaneous magnetic moment at the same temperature.

Additional observations of anomalous properties of $LiCu_2O_2$ around 150 K have been reported in electrophysical measurements [22]. The thermal evolution of the electrical conductivity shows peculiarities near 150 K, however with some frequency dependence of the temperature for the anomaly and some from sample to sample variation. The temperature dependence of the polarization current and dielectric permittivity exhibit non-monotonic dependence with drops in the 150 -160 K range. However, as yet no associated anomaly has been observed in heat capacity measurements on $LiCu_2O_2$ [6].

Interestingly, recent studies have shown that relatively small (from 1 to 10 %) replacements of Cu by Zn affect and even hinder the low temperature antiferromagnetic state of $LiCu_2O_2$ [23, 24]. A similar effect was observed in case of Ni doping [25]. Figure 5 shows a plot of the ZFC and FC magnetization along the $c$-axis vs. temperature in an applied field of 20 Oe for $LiCu_2O_2$ single crystals in which some copper has been substituted for silver: $LiCu_{1.94}Ag_{0.06}O_2$ and $LiCu_{1.86}Ag_{0.14}O_2$. A small amount of Ag preserves the spontaneous magnetic moment, whereas a slightly larger amount kills it; however, the antiferromagnetic ordering at low temperature (as reflected in the temperature dependence of the magnetization at low temperature) appears essentially unaffected by the doping.

**Conclusions**

The thermal evolution of lattice metrics of $LiCu_2O_2$ has been investigated by x-ray powder diffraction technique in the temperature range 12-295 K. An irregular lattice parameter behavior is found near $T_N$, confirming magnetoelastic-coupling effects. In addition, the weak spontaneous magnetic moment appearing at 150 K is accompanied by an anomalous temperature dependence of the thermal expansion along the *c*-axis around 150 K. Surprisingly, the orthorhombic distortion of LiCu$_2$O$_2$ increases with increasing temperature.

**Acknowledgements**

We thank the Swedish Research Council (VR), the Göran Gustafsson Foundation, and the Russian Foundation for Basic Research for financial support. We thank Rolf Berger for providing LiCu$_2$O$_2$ plate-like single crystals for comparison.

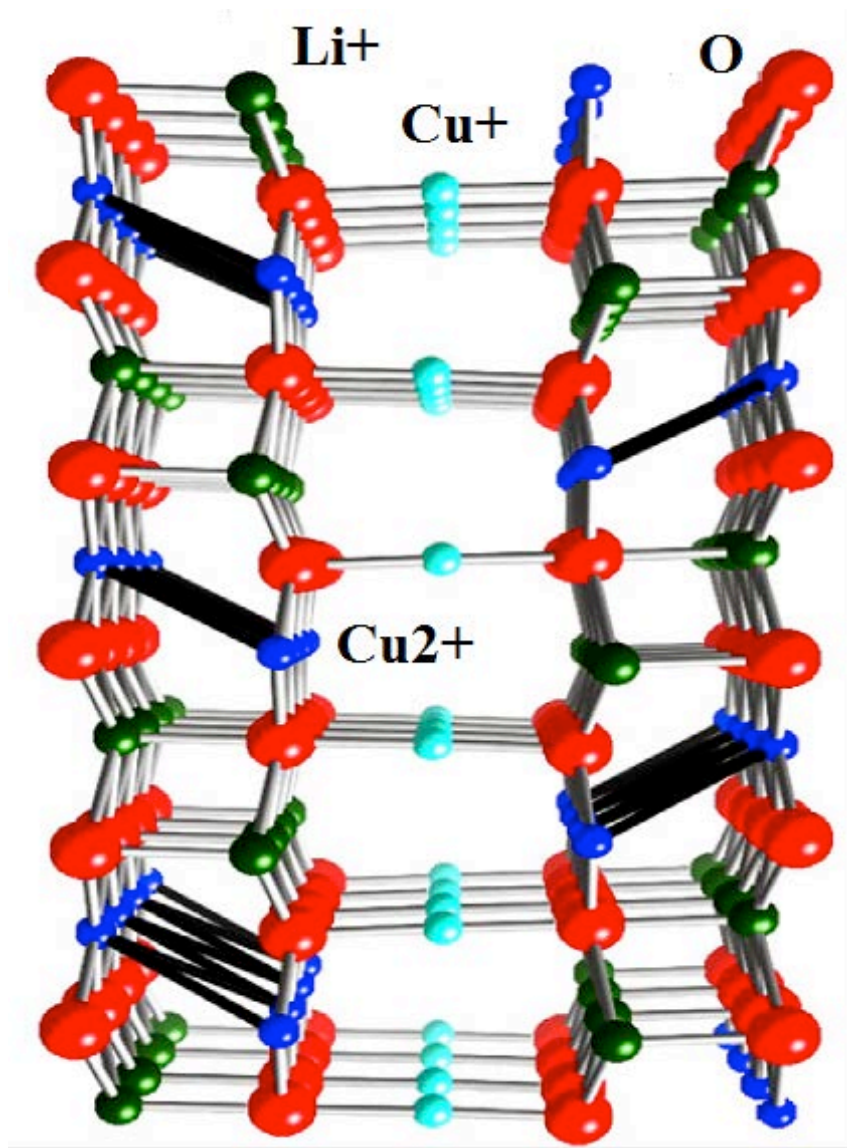

Figure 1 Projection of the crystal structure of LiCu$_2$O$_2$ along the *b*-direction. Zigzag Cu$^{2+}$ spin chains are presented by black chemical bonds.

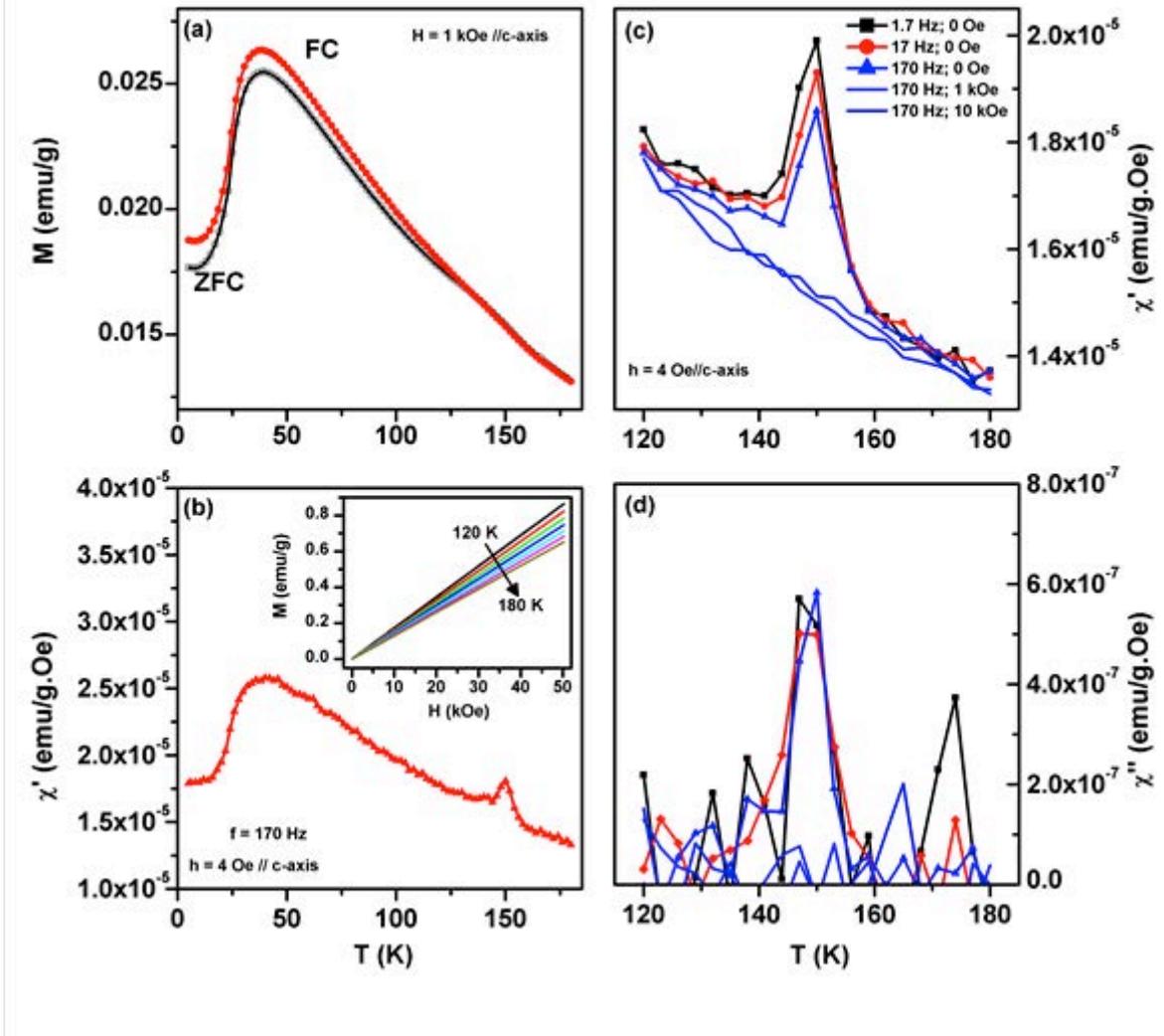

**Figure 2** (a) ZFC and FC magnetization measured with a relatively large field H (1 kOe) along the *c*-axis. (b) Temperature dependence of the in-phase component of the ac-susceptibilty $\chi'$. Frequency f =170 Hz and ac-excitation h = 4 Oe along the c-axis. The inset shows M vs. H curves at different temperatures. (c) Temperature dependence of $\chi'$ and (d) the out-of-phase component of the ac-susceptiblity $\chi''$ for three frequencies in absence of dc field (lines with markers), and collected while superimposing dc fields of 1 kOe and 10 kOe (simple lines, f = 170 Hz).

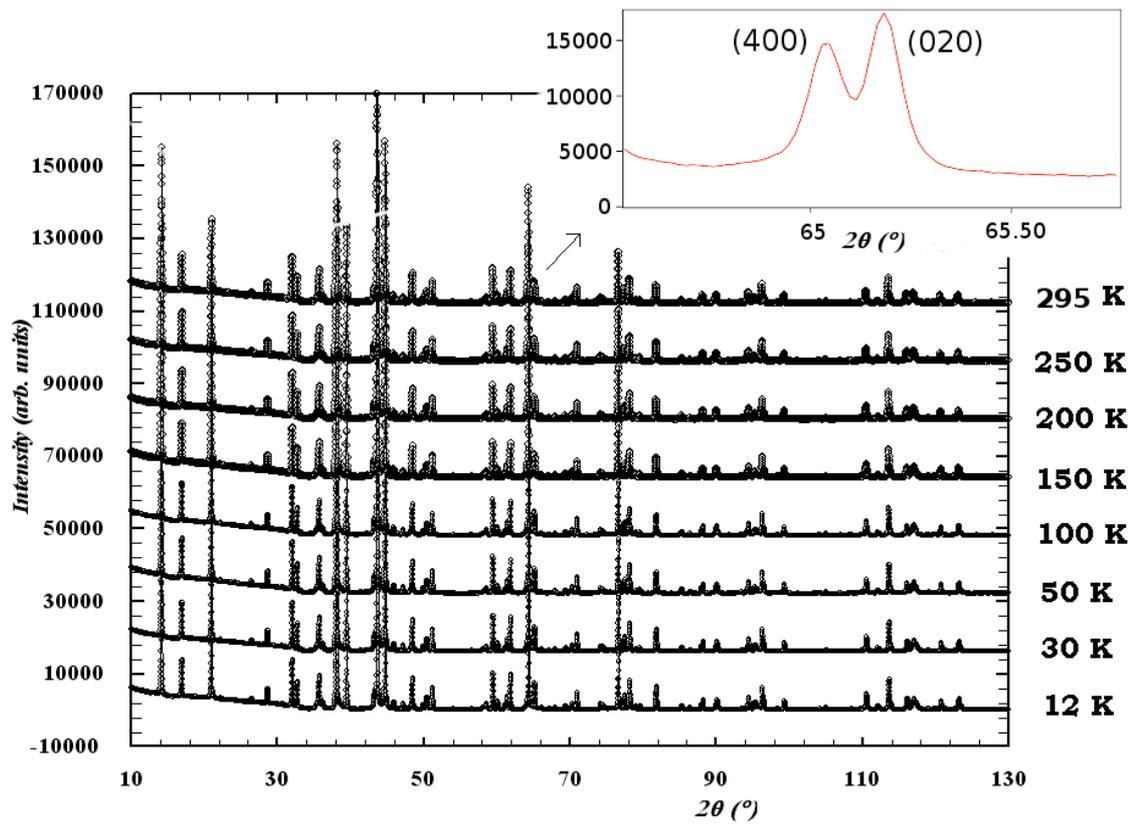

**Figure 3**. Variable temperature X-ray powder diffraction data for LiCu$_2$O$_2$
Powder patterns are shown from 12 K (bottom pattern) to 295 K (top pattern).
The inset shows a zoomed view of the reflections near 2θ ~ 65 degrees (T=295 K).

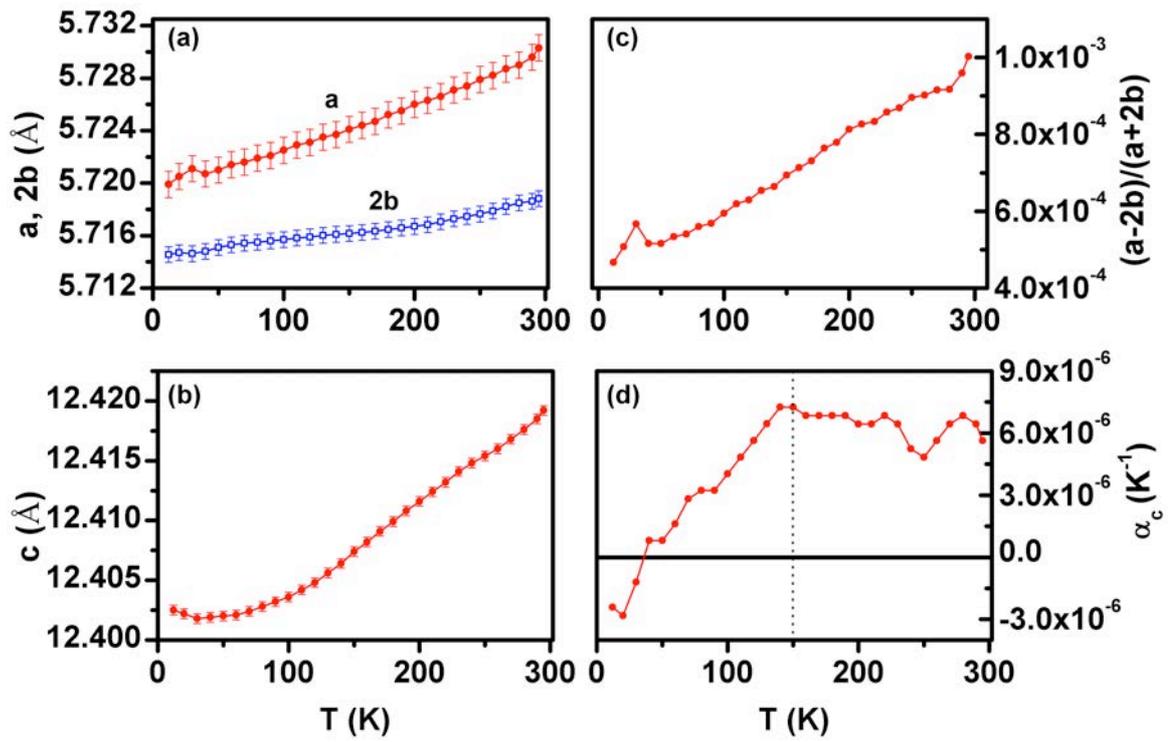

**Figure 4**. Left panels: thermal evolution of lattice parameters of LiCu$_2$O$_2$ in the temperature range 12 - 300 K. Right panels: (c) orthorhombic distortion and (d) thermal expansion coefficient along c direction of LiCu$_2$O$_2$ as a function of temperature.

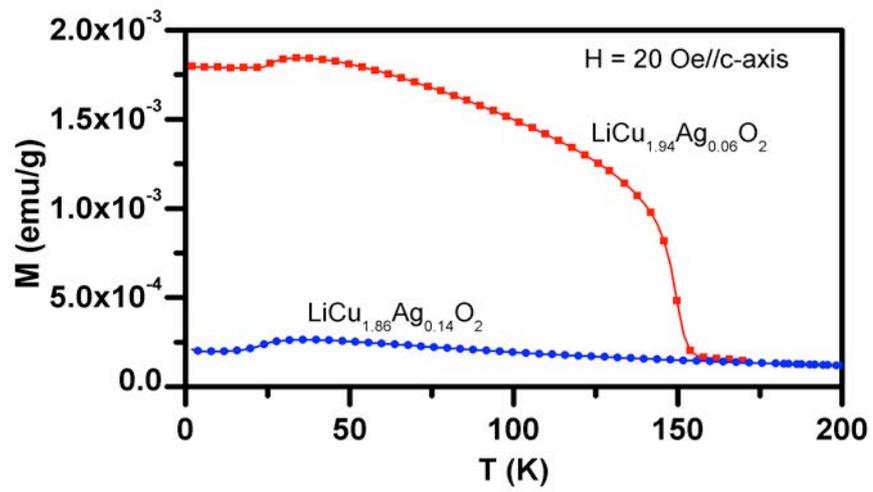

Figure 5. Temperature dependence of the FC magnetization for Ag-doped $LiCu_2O_2$ single-crystals, recorded in a small magnetic field H = 20 Oe applied along c-direction.